\documentclass[a4paper,11pt]{article}
\usepackage{jinstpub} 
\usepackage{siunitx}
\usepackage{lineno}

\title{\boldmath Test-beam results from MiniCACTUS-v2: A depleted monolithic CMOS timing sensor prototype}

\author[a]{Y.~Degerli} \note{Corresponding author.}
\author[a]{, R.~Aleksan}
\author[c,d]{, R.~Casanova}
\author[c,e]{, Y.~Gan}
\author[c,f]{, S.~Grinstein}
\author[a]{, F.~Guilloux}
\author[h]{, A.~Hanlon}
\author[g]{, T.~Hemperek}
\author[a]{, J.P.~Meyer}
\author[c]{, J.~Pinol}
\author[a,b]{, P.~Schwemling}
\author[h]{, E.~Vilella}

\affiliation[a]{IRFU, CEA, Universit\'{e} Paris-Saclay, F-91191 Gif-sur-Yvette, France}
\affiliation[b]{ Université Paris-Cité, Campus Grands Moulins, 75013, Paris, France}
\affiliation[c]{Institute for High Energy Physiscs (IFAE)
 Campus Autonomous University of Barcelona (UAB), 08193, Bellaterra (Barcelona), Spain}
 \affiliation[d]{Department of Microelectronics and Electronic Systems, Autonomous University of Barcelona (UAB),
 Campus Autonomous University of Barcelona (UAB), 08193, Bellaterra (Barcelona), Spain}
 \affiliation[f]{Catalan Institution for Research and Advanced Studies (ICREA)}
 \affiliation[g]{DECTRIS AG, Taefernweg 1, 5405 Baden-Daettwil, Switzerland}
 \affiliation[h]{Department of Physics, University of Liverpool,
Oliver Lodge Building, Oxford Street, Liverpool L69 7ZE, U.K}

\emailAdd{yavuz.degerli@cea.fr}

\abstract{MiniCACTUS-v2 is a monolithic sensor prototype designed in LF 150 nm CMOS process for time tagging of individual Minimum Ionizing Particles with an accuracy better than 100 ps. The sensing element is a deep n-well/p-substrate diode without internal amplification. To minimize detector capacitances, the analog front-ends and the discriminators for each pixel have been implemented outside the pixel, at the column level. After fabrication, the sensors have been thinned to \SI{150}{\micro m}, \SI{175}{\micro m} and \SI{200}{\micro m}  and then post-processed for backside biasing. The breakdown voltages measured on these sensors are higher than 500 V, ensuring the complete depletion of the charge collection volume. In this paper, we will focus on the time resolution measurements from a test-beam campaign conducted in July 2025 at SPS-CERN. During this period, several pixels from the 3 different sensor thicknesses have been tested at different bias voltages. The best time resolution measured is 48.88 ps on a 0.5 mm$\times$ 0.5 mm pixel from a \SI{175}{\micro m}-thick sensor at 500 V, with nominal settings for the on-chip analog front-end and discriminator.}


\begin{document}
\maketitle
\flushbottom

\section{Introduction}
\label{sec:intro}



Depleted monolithic timing detectors are currently being developed in the LF 150 nm CMOS process for future high energy experiments, either as possible post-phase 2 LHC upgrades or targeting at longer term projects like FCC-ee. The depleted CMOS technology is an interesting low-cost and reliable alternative to the current large-scale semiconductor timing detectors, largely based on Low Gain Avalanche Detectors (LGADs~\cite{lgad}). The radiation hardness of depleted CMOS sensor in this technology has already been demonstrated with tracking detectors developed for ATLAS Inner Tracker upgrades \cite{hv-cmos}. The first prototype designed in this process that meets the sub-100~ps timing results is MiniCACTUS-v1 \cite{minicactus_v1}. A time resolution of 65 ps has been measured at CERN-SPS with muons on a \SI{200}{\micro m}-thick sensor. While the results of this prototype were promising in terms of breakdown voltage and timing resolution, a significant coupling signals have been observed from the digital part to the sensor/analog part, causing ringing of the output signals \cite{minicactus_v1}. The origin of this problem has been attributed to the on-chip CMOS buffers driving millimeter long signal paths. Another requirement yet to be met was the long recovery time of the analog front-end, higher than the \SI{25}{ns} required for LHC applications. In order to address these issues, and also to have a sensor with larger homogeneous active area, a new prototype, called MiniCACTUS-v2, has been designed and fabricated in 2024 in the same process.

In this paper, experimental results from a test-beam campaign conducted in July 2025 at SPS-CERN on MiniCACTUS-v2 sensors with different thicknesses are presented. During this test-beam, we focused to measure the ultimate time resolution achievable with a sensor without internal amplification. 

\section{Sensor Description}

\begin{figure}[htbp]
\centering
\includegraphics[width=.5\textwidth]{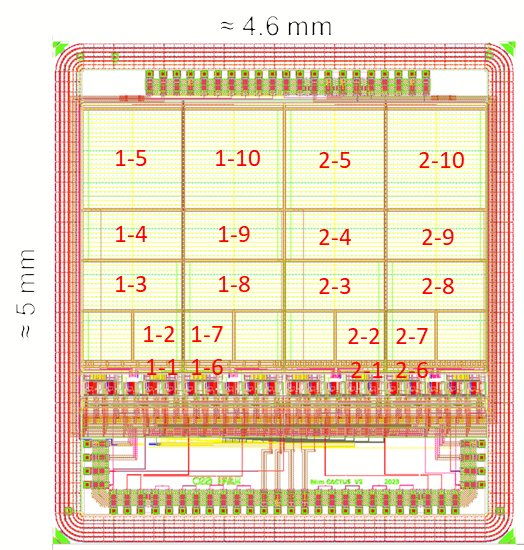}
\caption{Layout of the MiniCACTUS-v2 sensor.\label{layout}}
\end{figure}

The MiniCACTUS-v2 chip tested in this study is a monolithic CMOS sensor which includes an active array of 4 $\times$ 5 pixels, external guard-rings used to bias and deplete the high resistivity substrate ($\geq$ 2 k$\Omega$$\cdot$cm), an analog Front-End (FE) per pixel, a Slow Control (SC) interface and internal programmable biases through DACs. The dimensions of the chip are 5 mm $\times$ 4.6 mm. Its layout is shown in figure~\ref{layout}. 

The FE electronics consists of an AC-coupled preamplifier, a discriminator and a 5-bit DAC for threshold tuning. The AC coupling capacitances are integrated on the top of the charge collecting diodes. The FEs are implemented at the column-level, in order to avoid the large metal rails needed to distribute the power over the whole matrix which add significant parasitic capacitance to the detector capacitance for large diodes, as learn from the first CACTUS prototype developed for timing measurements \cite{cactus}. All DACs of the chip are programmable through a SC block. The baseline pixel pitches are 1 mm $\times$ 1 mm, 0.5 mm $\times$ 1 mm and 0.5 mm $\times$ 0.5 mm. There are also two small size pixels per column (\SI{50}{\micro m} $\times$ \SI{50}{\micro m} and \SI{50}{\micro m} $\times$ \SI{150}{\micro m}) implemented as test structures. For chosen pixels, the digital outputs of the discriminators and a buffered, slowed down copy of the preamplifier outputs are available on external pads. 

Several improvements have been done in this prototype to address the coupling issues from the digital signals described in the previous section: The distance between the outputs of the discriminators and output LVDS drivers has been shortened by moving the LVDS drivers from the pads to an area close to the discriminator outputs. Furthermore, these drivers have been placed inside a dedicated deep n-well. To shorten the Time over Threshold (ToT) of the discriminator output signals, 2 new preamplifier architectures have been designed and integrated into this chip. The first is an optimized Charge Sensitive Amplifier (CSA), the second a Voltage Pre-Amplifier (VPA), inspired from ALTIROC developments~\cite{9507972}. The 3 different amplifier architectures implemented in the chip have been detailed in \cite{minicactus_v2}. After fabrication, the standard High-Resistivity (HR) wafers have been thinned to 3 different thicknesses: \SI{150}{\micro m}, \SI{175}{\micro m} and \SI{200}{\micro m}. Then, they have been post-processed for backside polarization (boron implant, thermal annealing, metalization), essential for homogeneous charge collection in timing applications. It is worth noting that to compute the effective active sensor thicknesses, one should remove approximately \SI{15}{\micro m} from these values due to the top metalization layers of the standard CMOS process.

\section{In-Lab Tests}

The post-processed devices demonstrate significantly improved performance with respect to the previous sensor version, MiniCACTUS-v1.
In particular, the mitigation of digital-to-analog coupling issues was confirmed through in-lab characterization, validating the effectiveness of the implemented design improvements.

The sensor performance was first evaluated under beam conditions at CERN-SPS in July 2024, yielding a best timing resolution of 59.9~ps for a 0.5~mm $\times$ 0.5~mm pixel from a \SI{175}{\micro m} thick sensor operated at 350~V bias \cite{minicactus_v2_tb1}. These results demonstrate the intrinsic timing capabilities of the device architecture. Performance was however limited by high voltage problems, with devices randomly failing even at relatively moderate high voltage.

High and unstable leakage currents (> 2-3\,µA) were traced to contamination originating from a no-clean PCB manufacturing process and were effectively suppressed by ultrasonic cleaning in isopropyl alcohol. Electrical discharges near the high-voltage interface were eliminated by redesigning the PCB layout to increase the separation between the high-voltage strip and grounded components. The high voltage connector has also been relocated on the PCB.
In addition, unexpected device failures were prevented by leaving optional bias pads located above the chip guard rings unbonded, thereby avoiding detrimental electric field configurations.

\begin{figure}[htbp]
\includegraphics[width=.46\textwidth]{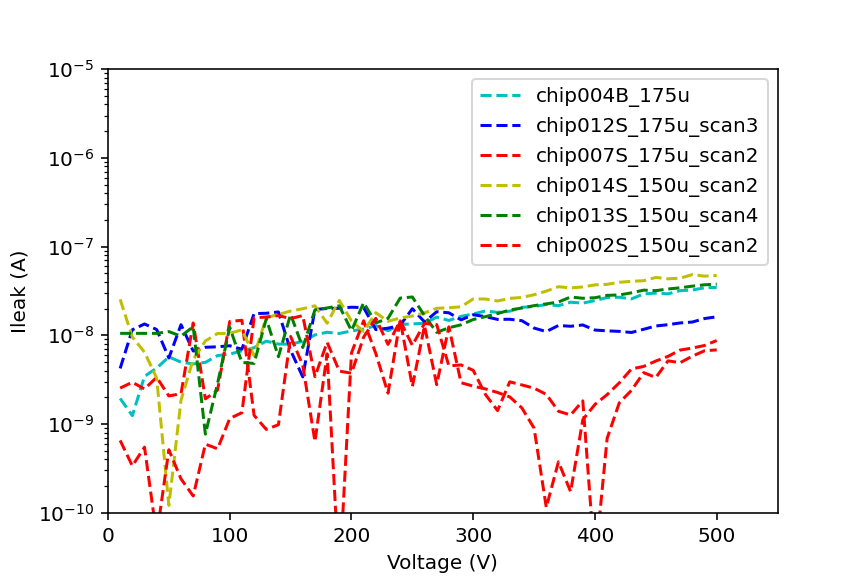}
\qquad
\includegraphics[width=.46\textwidth]{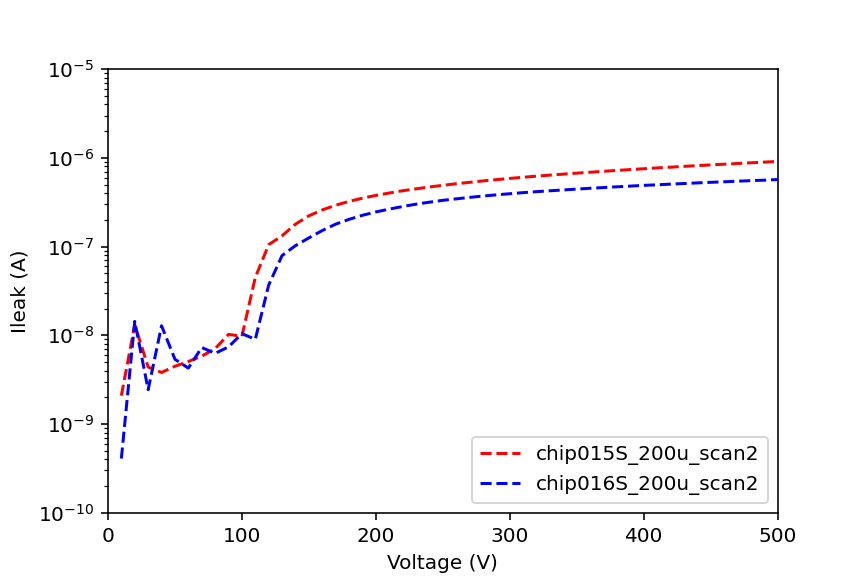}
\caption{I-V curves measured on several \SI{150}{\micro m},  \SI{175}{\micro m} (left) and \SI{200}{\micro m} (right) sensors. All tested sensors can be safely biased up to 500 V.\label{i-v}}
\end{figure}

Following these improvements, the electrical performance of the sensors was reassessed. The I--V characteristics of several \SI{150}{\micro m} and \SI{175}{\micro m} thick sensors measured up to 500~V are shown in left side of figure~\ref{i-v}. The leakage currents are extremely low over the full bias range, frequently falling below the resolution limit of the Keithley~2470 source-measure unit used for simultaneous biasing and current measurement. The observed fluctuations are therefore attributed to the noise floor of the measurement setup. Sensors with a thickness of \SI{200}{\micro m}, processed on the same wafers and sharing an identical design, exhibit similarly high breakdown voltages exceeding 500~V, albeit with higher leakage currents (right side of figure~\ref{i-v}). While the origin of this difference remains under investigation, it is attributed to variations introduced during post-processing. All sensor variants can nonetheless be safely operated at bias voltages up to 500~V.

\section{Test-beam Setup}
\begin{figure}[htbp]
\centering
\includegraphics[width=.45\textwidth]{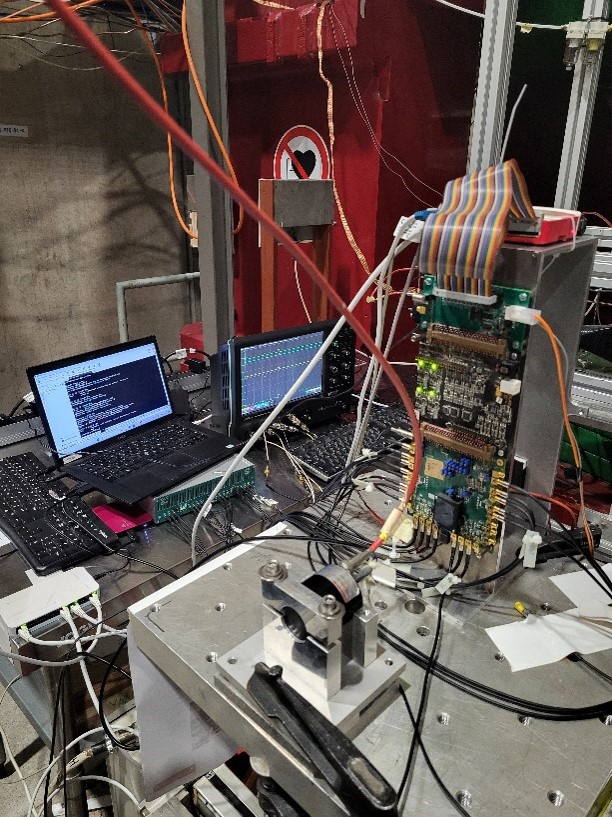}
\caption{Picture of the test-beam setup showing the the sensor under test, the PCBs, the PMTs, and the scope used for data acquisition.\label{test-beam}}
\end{figure}

The new test-beam campaign has taken place at SPS (CERN, North Area H4 beamline) during the gaseous detector R\&D (DRD1) test-beam period with 180 GeV/c muons in July 2025. Figure~\ref{test-beam} shows the picture of the test-beam setup including, among others, the sensor under test, the PCBs, the Photo-Multiplier Tubes (PMTs) used as timing references, and the digitizing scope. The power supplies are not visible in this picture.

The data acquisition has been done with a 12 bit-digitizing scope (LeCroy HDO4104A, 4 channels, 1 GHz bandwidth). 2 Hamamatsu H11934-100 PMTs have been used as time references. The photocathodes of the PMTs are coupled to a NE110 organic scintillator with dimensions 1.5~cm~$\times$~1.5~cm  and 0.5 cm-thickness, located on the center of the photocathode. PMT analog signals were readout by the scope through a 15 dB SMA attenuator connected to a $50\,\Omega$ cable. The attenuator was needed to avoid saturating the scope input.

Several pixels from different sensors with different thicknesses have been tested at various bias voltages up to 500 V. During this test-beam campaign, since the main objective was the measurement of the ultimate time resolution achievable with a sensor without internal amplification, we started the tests on a 0.5 mm $\times$ 0.5 mm pixel that gave the best result during the previous test-beam, and then scanned this pixel at different sensor bias voltages for different device thicknesses. At the end of the test-beam, a reduced data set has also been taken on 0.5 mm $\times$ 1.0 mm pixels.

\section{Data Analysis and Results}

During the data analysis, signal base-line positions and resolutions are computed event by event using the pre-pulse points of the digitized signals.

\begin{figure}[htbp]
\includegraphics[width=.45\textwidth]{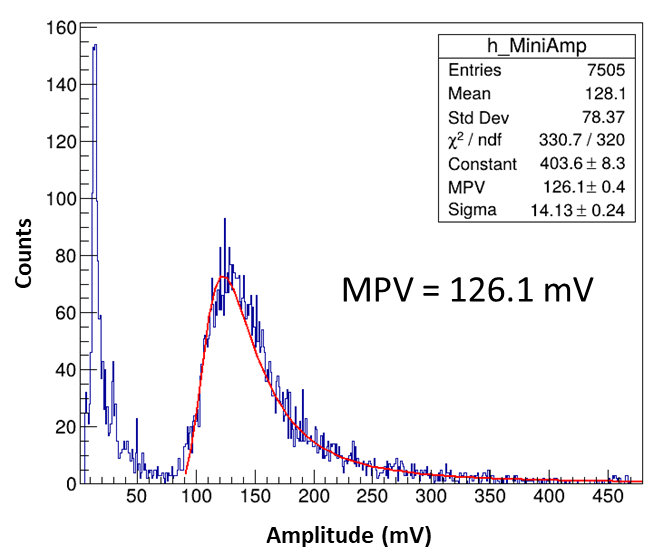}
\qquad
\includegraphics[width=.48\textwidth]{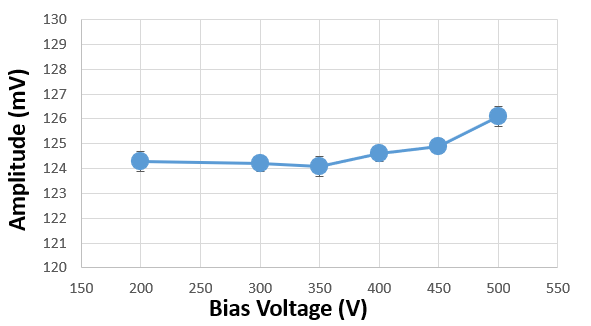}
\caption{Measured (a) amplitude distribution on the analog monitoring output (AmpOut) with MIPs at 500 V bias voltage, (b) MPV as a function of sensor bias voltage showing very small variation in this range (0.5~mm~$\times~$~1.0~mm pixel from a 175 µm-thick sensor).\label{mpv}}
\end{figure}

The amplitude distribution measured on the analog monitoring output (AmpOut) at 500~V is shown in figure~\ref{mpv} for a 0.5~mm~$\times~$~1.0~mm pixel from the \SI{175}{\micro m}-thick sensor, together with the maximum probable value signal amplitude  as a function of the sensor bias voltage. Since the sensor is already depleted at these voltages, the amplitude variation is very small over the applied voltage range. The amplitude information is also used event by event for Time Walk (TW) corrections.

\begin{figure}[htbp]
\centering
\includegraphics[width=.8\textwidth]{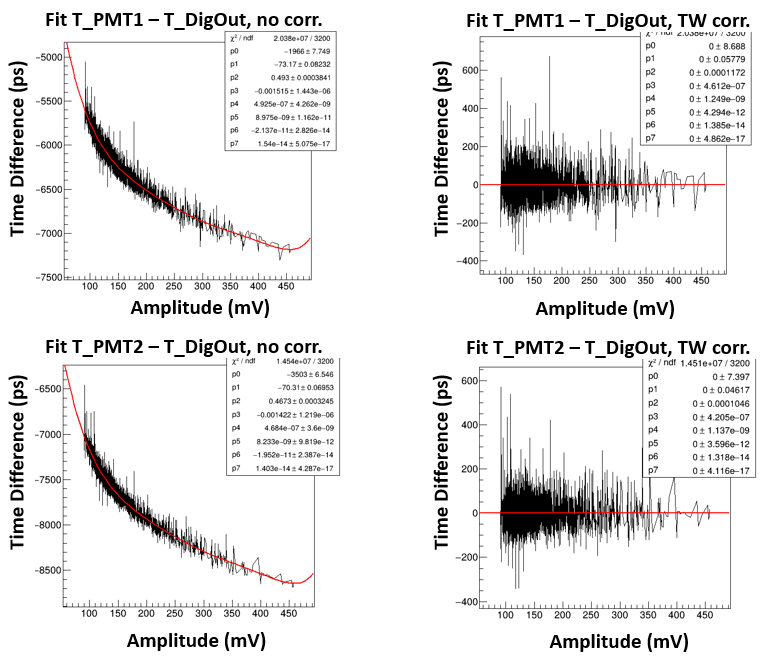}
\caption{Measured time difference distributions between the DUT digital output (DigOut) and the 2 PMTs (PMT1 and PMT2) as a function of analog signal amplitude (AmpOut) for TW corrections.\label{tw-corrections}}
\end{figure}

PMT signals are relatively slow, with a risetime of a few ns,  compared to MiniCACTUS-v2 digital output, nevertheless the time information is contained into the rise up of the pulse (first photo-electron). In order to filter noise in the very beginning part of the PMT signal, we do an off-line constant fraction discrimination at 10\% level of the signal maximum and a polynomial fit selecting points around the discriminator level. Both the polynomial function order and the number of points in the fit have been varied in order to minimize the dispersion of the two PMT time difference, i.e. the best time resolution. This is obtained with 14 points, representing 1.4~ns,  and a second order polynomial function. An offline 50\% constant fraction threshold has been applied to the MiniCACTUS-v2 digital signal. As for the PMT signals, points selected around the discriminator level have been fitted with a polynomial function. For the MiniCACTUS-v2 analog signal amplitude, we keep the maximum value above the base-line. The left side of figure~\ref{tw-corrections} shows the time differences between the MiniCACTUS-v2 digital signal (DigOut) and the PMT signal rising edge time as a function of the MiniCACTUS-v2 analog signal amplitude (AmpOut) for both PMT’s. A 7th order polynomial fit is applied to correct for this TW. The right side of figure~\ref{tw-corrections} shows the same plots after application of the TW correction. The time resolution of the PMTS’s and MiniCACTUS-v2 is obtained by inverting the matrix of the three times differences. Again, the number of points and the polynomial order for the MiniCACTUS-v2 digital time reconstruction has been varied to find the best resolution. This is obtained with a third order polynomial function fitted over 6 points, corresponding to 600~ps, around the 50\% level threshold. The pulse reconstruction procedure has been optimized on the first good run of the testbeam period and then used without any parameter change for the analysis of all the data. The upper plots of figure~\ref{tw-corrections} are for PMT1-DigOut and lower ones for PMT2-DigOut.

\begin{figure}[htbp]
\centering
\includegraphics[width=1.0\textwidth]{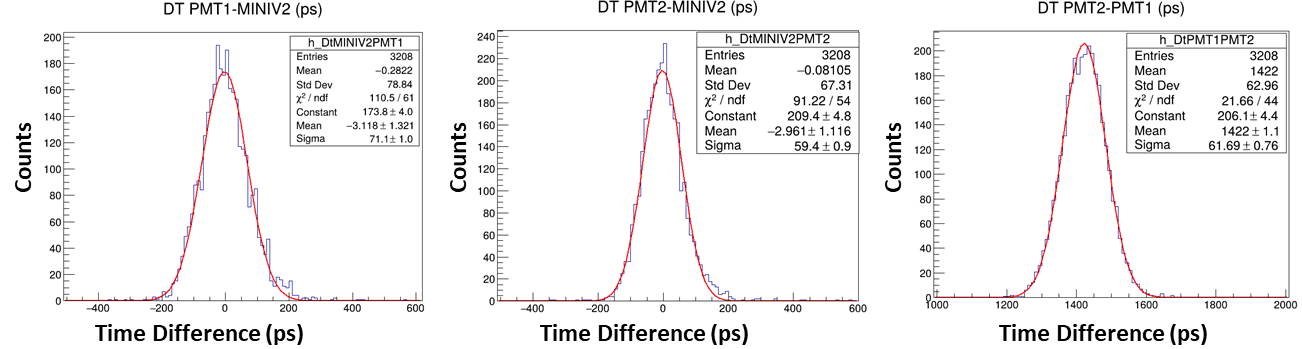}
\caption{Measured 3 time difference distributions (PMT1-DUT, PMT2-DUT and PMT1-PMT2). The DUT is MiniCACTUS-v2. After time walk correction, the extracted time resolution for the MiniCACTUS-v2 is 48.88 ps (0.5 mm $\times$ 0.5 mm pixel from a 175 µm-thick sensor, 500 V bias voltage).\label{timings}}
\end{figure}

Figure~\ref{timings} shows the time difference distributions between the 3 devices after TW corrections. The results are also summarized in table~\ref{measured-sigmas}. By inverting the matrix, one can extract the individual time resolutions of each device given in table~\ref{extracted-sigmas}. For this measurement, the extracted value for the DUT (MiniCACTUS-v2) is 48.88 ps.

\begin{table}[htbp]
\centering
\caption{Measured Sigma matrix (time resolutions) for MiniCACTUS-v2, PMT1 and PMT2 after time walk corrections.\label{measured-sigmas}}
\smallskip
\begin{tabular}{l|r c}
\hline
Devices & Resolution (ps) & error (ps)\\
\hline
MiniCACTUS-PMT1 & 71.10 & 1.00\\
MiniCACTUS-PMT2 & 59.40 & 0.90\\
PMT2-PMT1 & 61.69 & 0.76\\
\hline
\end{tabular}
\end{table}

\begin{table}[htbp]
\centering
\caption{Extracted time resolution of each device after matrix inversion (0.5 mm $\times$ 0.5 mm pixel from a 175 µm-thick sensor at 500 V bias voltage).\label{extracted-sigmas}}
\smallskip
\begin{tabular}{l|r c}
\hline
Devices & Resolution (ps) & error (ps)\\
\hline
\textbf{MiniCACTUS} & \textbf{48.88} & 1.03\\
PMT1 & 51.64 & 0.97\\
PMT2 & 33.76 & 1.49\\
\hline
\end{tabular}
\end{table}

\begin{figure}[htbp]
\centering
\includegraphics[width=.45\textwidth]{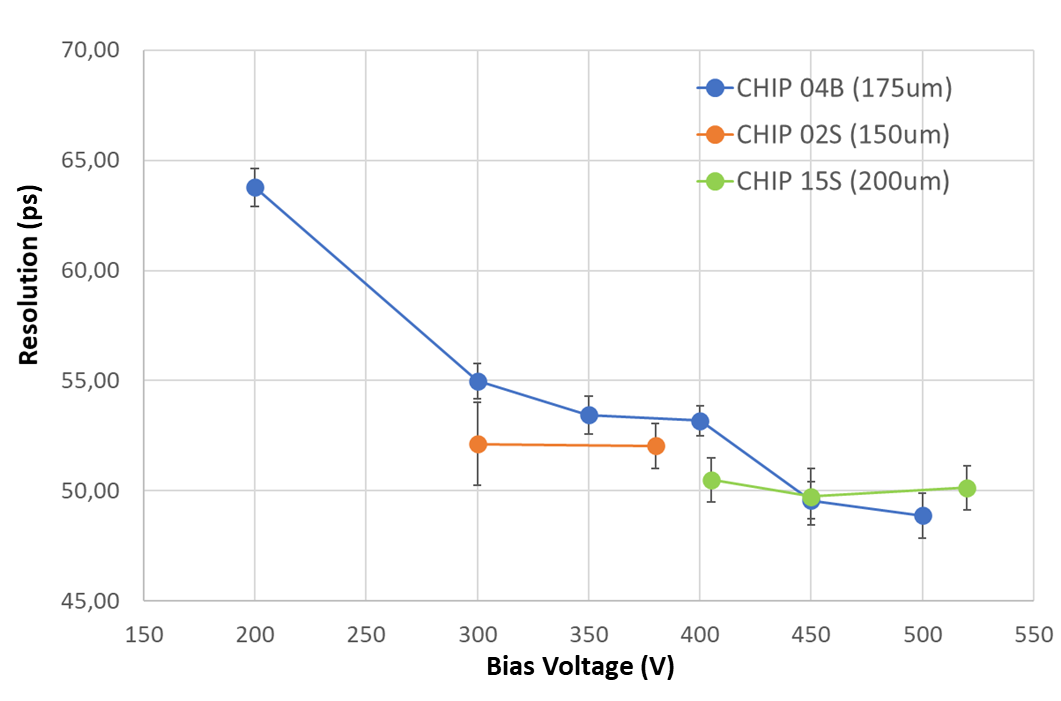}
\qquad
\includegraphics[width=.48\textwidth]{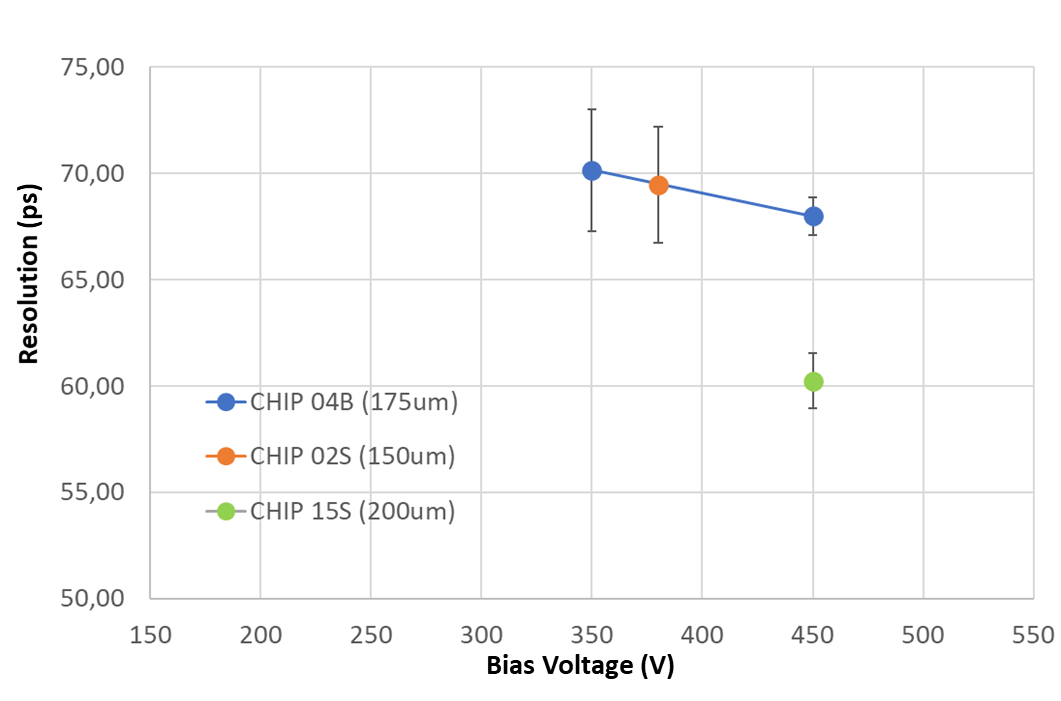}
\caption{Summary of time resolution measurements from 3 sensors with different thicknesses as a function of bias sensor voltage for (a) 0.5 mm $\times$ 0.5 mm pixels and (b)  0.5 mm $\times$ 1.0 mm pixels.\label{timing-pix12_13}}
\end{figure} 

The time resolution measured on a 0.5 mm $\times$ 0.5 mm pixel (pixel 1-2) as a function of the sensor bias voltage is shown in figure~\ref{timing-pix12_13}a for 3 different thicknesses. At high voltage values, the results of for \SI{175}{\micro m} and \SI{200}{\micro m} sensor thicknesses are quite close. 

Figure~\ref{timing-pix12_13}b summarizes the measurements on a 0.5 mm $\times$ 1.0 mm pixel (pixel 1-3). The best resolution measured is on the \SI{200}{\micro m} sensor thickness for this pixel. 




\section{Conclusions}

Several MiniCACTUS-v2 sensors designed in LF 150 nm process with 3 different thicknesses have been tested extensively during 2 weeks with MIPs at CERN-SPS during July 2025. Up to now, the best timing result obtained is 48.88 ps on a 0.5 mm × 0.5 mm pixel from a \SI{175}{\micro m} wafer, including the on-chip analog front-end and the discriminator.
Other thicknesses tested in this test-beam, \SI{200}{\micro m} and \SI{150}{\micro m}, give comparable results. Especially, for the 0.5 mm $\times$ 0.5 mm pixel from \SI{200}{\micro m} wafer the measured timing resolution at high bias voltages is $\sim$50 ps. For the 0.5 mm $\times$ 1.0 mm pixels, so far the best timing obtained is 60.23 ps, with a limited testing time that did not allow an extensive study. These results obtained with a depleted monolithic sensor without internal amplification validated the sub-100ps requirement and pave the way for a full monolithic sensor embedding a time to digital converter and data processing units for future high energy applications.

\acknowledgments
This project has received funding from the European Union's Innovation programme under grant no 101004761 (AIDAInnova). It has also received funding from the P2I department of Université Paris-Saclay, and was also partially supported by MICIIN (Spain) with funding from European Union NextGenerationEU(PRTR-C17.I1), and from the Generalitat de Catalunya.

The authors would finally like to acknowledge the DRD1 team for their kindful help during the test-beam campaign at SPS-CERN. 




\end{document}